\documentclass[a4paper]{jpconf}
\usepackage[dvipdfmx]{graphicx}
\usepackage{latexsym}
\usepackage{amsmath}
\usepackage{braket}

\begin{document}
%\title{Calibration for liquid noble gas detectors using $^{220}$Rn}
\title{Using $^{220}$Rn to calibrate liquid noble gas detectors}
\author{M~Kobayashi, M~Yamashita, A~Takeda, K~Kishimoto and S~Moriyama}

\address{Kamioka Observatory, Institute for Cosmic Ray Research, the University
of Tokyo, Higashi-Mozumi, Kamioka, Hida, Gifu, 506-1205, Japan}

\ead{mkoba@km.icrr.u-tokyo.ac.jp}

\begin{abstract}
In this paper, we describe $^{220}$Rn calibration source that was developed for liquid noble gas detectors.
The key advantage of this source is that it can provide $^{212}$Bi--$^{212}$Po consecutive events,
which enables us to evaluate the vertex resolution of a detector at low energy by comparing low-energy events of $^{212}$Bi 
and corresponding higher-energy $\alpha$-rays from $^{212}$Po. 
Since $^{220}$Rn is a noble gas, a hot metal getter can be used when introduced using xenon as the carrier gas.
In addition, no long-life radioactive isotopes are left behind in the detector after the calibration is complete; this has clear advantage over the use of $^{222}$Rn which leaves long-life radioactivity, i.e., $^{210}$Pb. 
Using a small liquid xenon test chamber, we developed a system to introduce $^{220}$Rn via the xenon carrier gas; we demonstrated the successful introduction of $6\times 10^2$ $^{220}$Rn atoms in our test environment.
\end{abstract}

\section{Introduction}
In recent years liquid noble gases, such as liquid xenon (LXe) or liquid argon (LAr), have been increasingly used for particle detectors. 
Many sensitive searches for dark matter particles, axions, and particles beyond the standard model were conducted using such LXe and LAr detectors \cite{lux, XENON100, DS50, xmass}.
These studies require a sufficient understanding of detector response from $\sim$O(100)keV down to the detector threshold caused by nuclear and electron recoils.
In particular, an energy calibration and evaluation of vertex reconstruction of the detector are an important means to interpreting experimental results.
To evaluate the energy response of a detector, neutron sources and small accelerators are commonly used for nuclear recoil events, 
whereas the direct introduction of radio-active sources (e.g.,$^{\mathrm{83m}}$Kr and tritiated ethane) into detectors is typically performed via electron recoil events.
Conversely, methods for evaluating reconstruction performance of vertex determination for low-energy events region are not well-established, because true vertices of diffused radioactive decay are not given independently.
To overcome this problem, Kim el al.~proposed the use of low-energy events whose event vertices are accurately determined by accompanying higher-energy events \cite{NDKim}.
In this paper, we report on the development of a calibration source based on $^{220}$Rn for liquid-noble-gas detectors.\\
Here, $^{220}$Rn is one of the isotopes of Rn located at the middle of the Thorium series.
There are several advantages of using $^{220}$Rn calibration as a source, including the following:
\begin{enumerate}

\item  Presence of $^{212}$Bi--$^{212}$Po consecutive decays with short life time of $^{212}$Po (i.e., $\tau_{\frac{1}{2}}$ = 299 ns) \\
Both $^{212}$Bi and $^{212}$Po are daughter nuclei of $^{220}$Rn and provide $\beta$-$\alpha$ consecutive decays.
Since the lifetime of $^{212}$Po is short, the difference between decay vertices of  $^{212}$Po and $^{212}$Bi is negligible, though they freely move in a liquid medium.
Further, $^{212}$Po provide an accurate determination of the decay vertex by reconstruction because of a large number of scintillation photons. 
Thus enabling us to use the difference of reconstructed vertices of low-energy $\beta$ and $\alpha$-rays to evaluate the performance of reconstruction at low energy.
Since they are expected to uniformly distribute in the detector, an elevation of position dependence of vertex resolution is possible as well.

\item  No long-life radioactivity downstream of $^{220}$Rn \\
The second advantage is that there are no long-life radioisotopes downstream of $^{220}$Rn.
Because of this, radioactivity that is introduced can decay out rather quickly.
Though $^{212}$Bi and $^{212}$Po in a Uranium chain also cause consecutive decays with a half-life of 164 $\mathrm{\mu}$s,
decaying out of all radioactivity takes a longer amount of time; more specifically, $^{222}$Rn and $^{210}$Pb have half-lives of 3.8 days and 20 years, respectively,
which is undesirable since it will cause an increase in background levels.

\item  Existence of $^{220}$Rn as a noble gas \\
The third advantage is that $^{220}$Rn is a noble gas, which is useful for suppressing other impurities since $^{220}$Rn can go through a hot metal getter typically used to purify a noble gas.
Here, the getter removes impurities such as O$_2$, H$_2$O, and oily molecules that may impact the performance of the detector. 
\end{enumerate}

\section{Proposed system for calibrating via $^{220}$Rn}
\subsection{Source}

We first note that there are publicly available thorium sources. In our calibration system, we used 50 pieces of lantern mantle containing thorium, as shown in Fig.\ \ref{LM}, CAPTAIN STAG M-7911.
Thorium is useful for generating strong light. Direct measurements using a Ge detector show $^{220}$Rn activity with 1.2 kBq per one piece of lantern mantle.
Emanating gas from some of these pieces of lantern mantle was measured and found to contain $\sim$30 Bq of $^{220}$Rn per one piece when in radioactive-decay equilibrium.  

\subsection{Detector and $^{220}$Rn injection system}
\begin{figure}[h]
\begin{minipage}{0.6\columnwidth}
\centering
\includegraphics[clip, width=7cm]{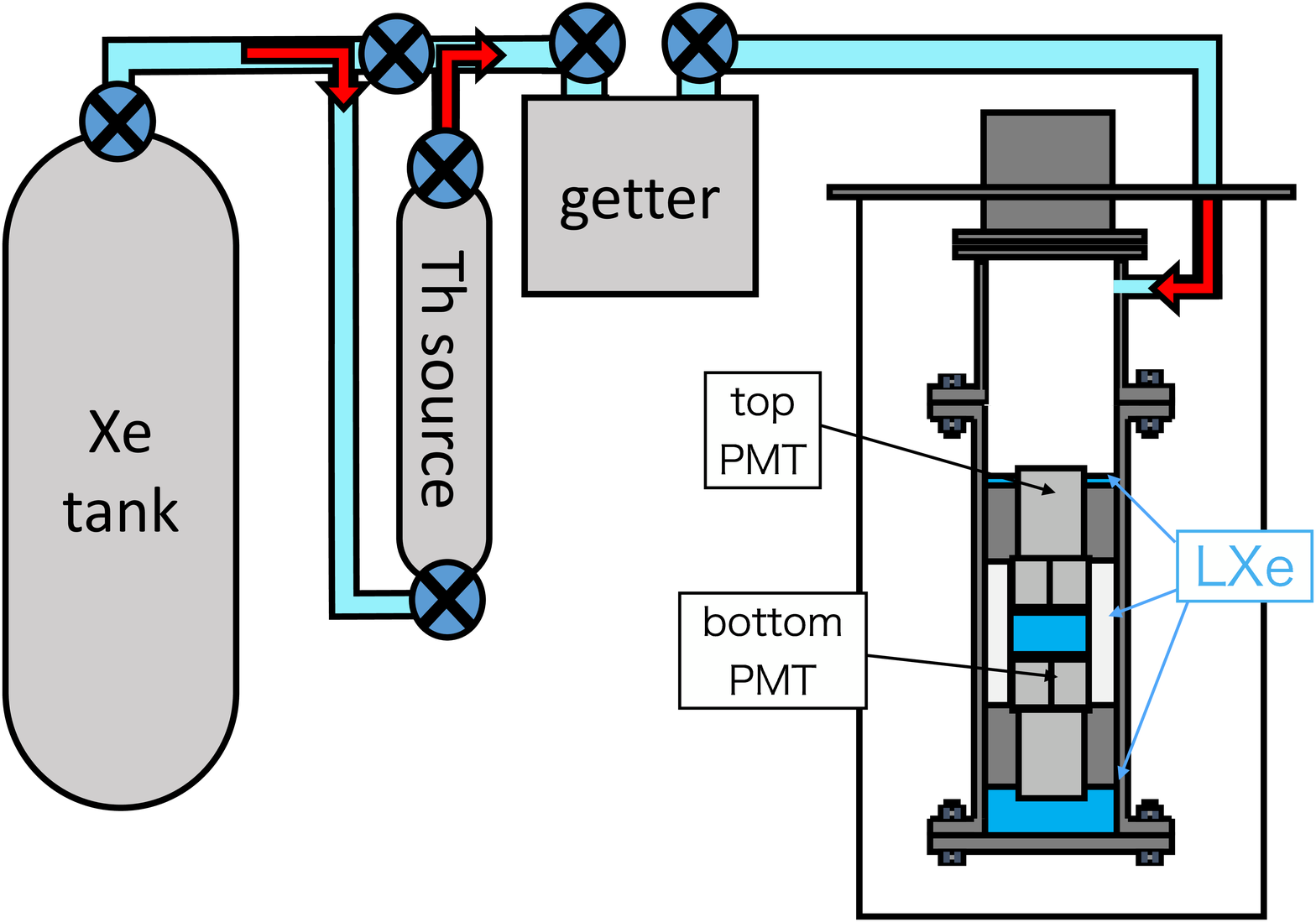}
\caption{Overview of the detector and gas line, with red arrows showing the introduction of the flow of $^{220}$Rn with Xe.} 
\label{Xeline}
\end{minipage}
\hspace{5mm}
\begin{minipage}{0.3\columnwidth}
\centering
\includegraphics[clip, width=3cm]{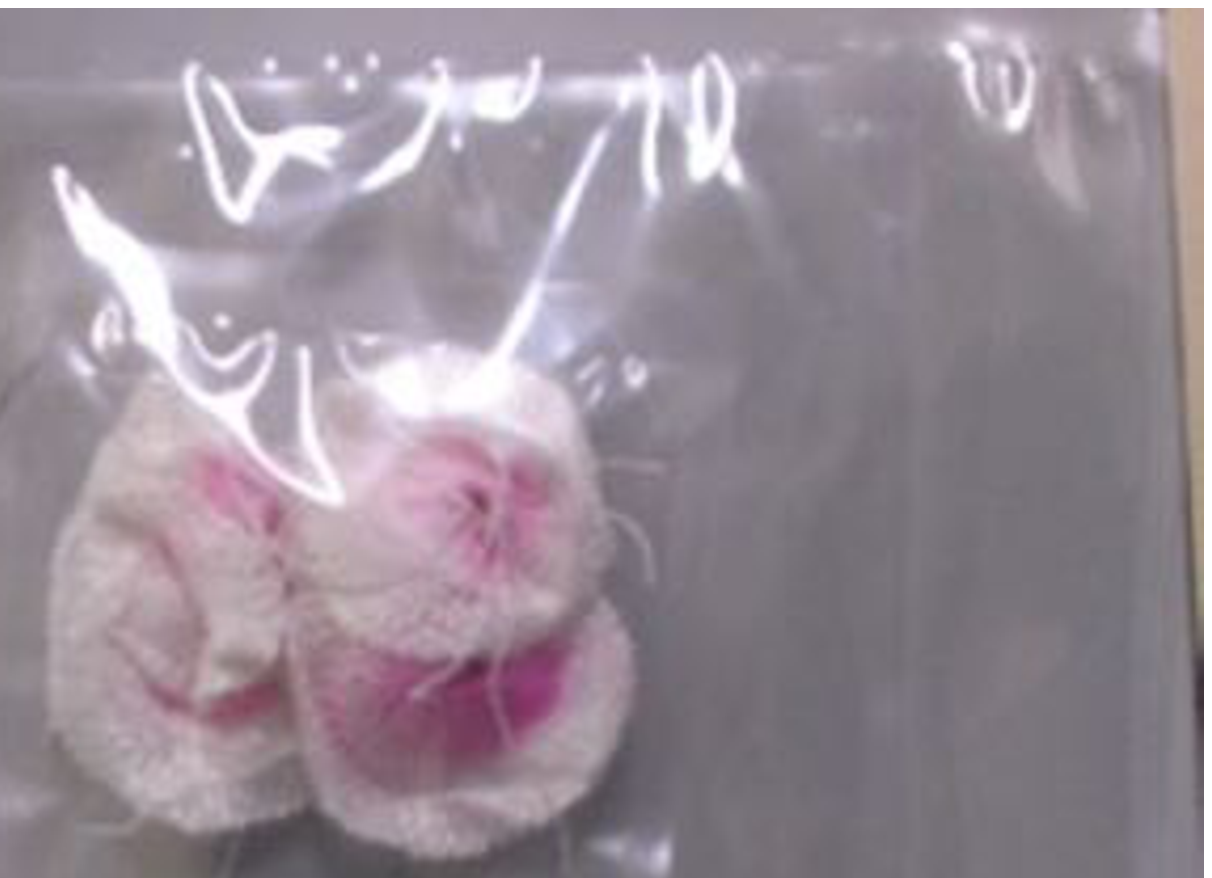}
\caption{Lantern mantle (CAPTAIN STAG M-7911) used as our thorium source; overall, 50 pieces of mantle were used. } 
\label{LM}
\end{minipage}
\end{figure}

The key issue to introduce $^{220}$Rn is its short half-life of 56 s.
For our system, we used a small LXe chamber located in the Kamioka mine.
Its inner cylindrical chamber was filled by LXe while its outer chamber was used for vacuum insulation.
The detector used 1.7 kg LXe and two photomultiplier tubes (PMTs) HAMAMATSU R10789;
the PMTs were set face-to-face using an aluminum holder and a polytetrafluoroethylene (PTFE) cylinder, as illustrated in Fig.\ \ref{Xeline}.
The detector was connected to a large Xe gas bottle containing $\sim$6 kg of xenon gas, with xenon introduced into the chamber from the bottle. 
The gaseous xenon were liquefied and refrigerated using temperature feedback;
here, its cold head was located above the top of the PMTs.

A container with a thorium source emanating $^{220}$Rn was then connected between the gas bottle and LXe chamber.
The diameter of the source container was 4 cm, and its length was 40 cm; 
the container had a volume of 502 $\mathrm{cm}^3$.
The 50 pieces of lantern mantle were kept in the source container. 
To introduce $^{220}$Rn gas, xenon gas from the gas bottle was used as the carrier. 
The xenon carrier gas introduced the emanated $^{220}$Rn gas into the detector through a getter. 
Here, the getter was necessary to remove impurities from the pieces of lantern mantle.
At the cold head, the mixture of carrier gas and $^{220}$Rn gas was liquefied.
To avoid a sudden change of pressure or temperature of the detector, the flow rate of the gas mixture was set to two liters per minute and maintained for two min.
In total, four liters of xenon carrier gas were introduced with $^{220}$Rn.
This quick operation is important for realizing the successful introduction of $^{220}$Rn.    

\subsection{Data acquisition trigger and energy calibration for LXe chamber}
The detector was equipped with a flash analog-to-digital converter (FADC) with 1 ns sampling and a 1 V dynamic range. 
The output of the two PMTs was amplified and fed into the FADC. Another output of the amplified signal was used to make a coincidence signal to trigger data acquisition.
The discrimination for each PMT signal was $\sim$30 keV, and coincidence was issued when two discriminator signals existed within 150 ns of each other.

\begin{figure}[h]
\begin{minipage}{0.5\columnwidth}
\centering
\includegraphics[clip,width=6cm]{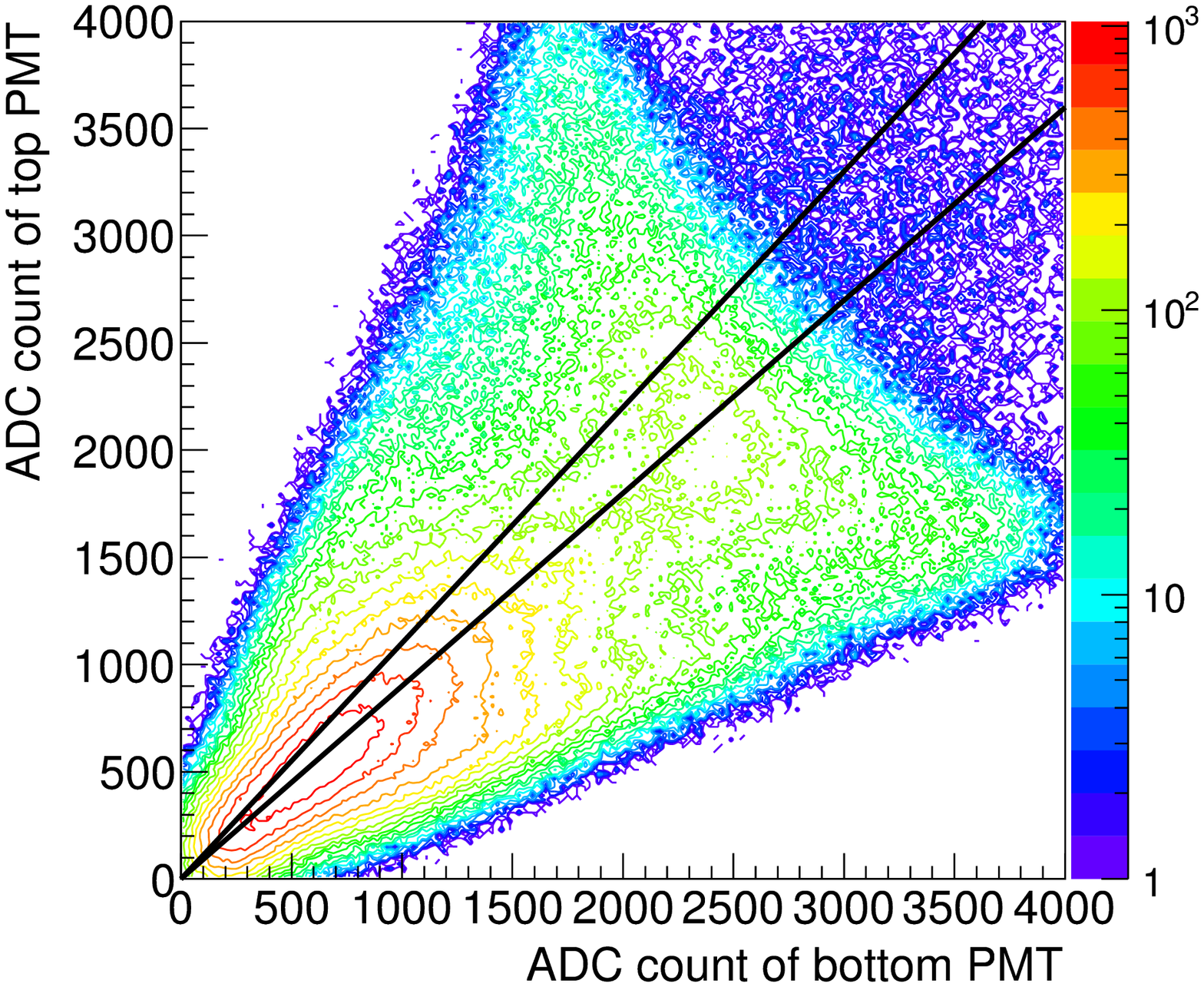}
\caption{Contour plot of observed events with a $^{137}$Cs source; here, the data shown within the black line (i.e., 0.9 $<$ Top/Bottom $<$ 1.1) are shown in the red histograms of Fig.\ 4. } 
\label{Cscalib1}
\end{minipage}
\hspace{5mm}
\begin{minipage}{0.5\columnwidth}
\centering
\includegraphics[clip,width=5.5cm]{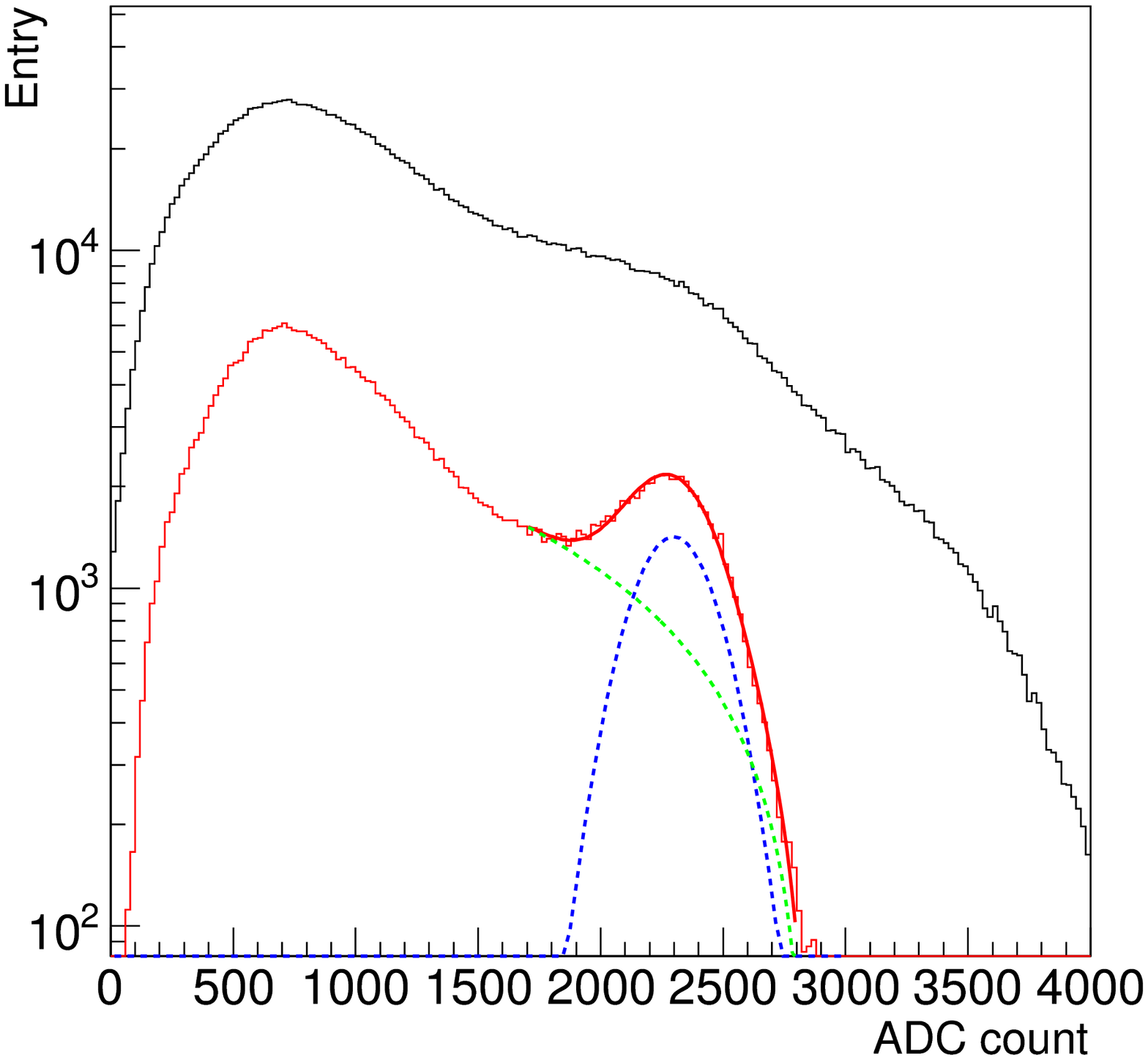}
\caption{ADC count distribution with a $^{137}$Cs source; here, the black histogram shows all events, whereas the red histogram shows events within the black lines of Fig.\ 3; red, blue, and green curves depict the fitting function, Gaussian component, and linear component, respectively.} 
\label{Cscalib2}
\end{minipage}
\end{figure}

To calibrate an energy scale of the detector and monitor its stability, we used a $^{137}$Cs external $\gamma$-ray source. 
This source was located at the bottom of the detector, outside of the outer chamber.
Because there is position dependence of the light yield in the inner chamber, events happening around the center were extracted based on the balance of observed photoelectrons from the two PMTs to form a photoelectric peak caused by 667 keV $\gamma$-rays, as shown in Fig.\ \ref{Cscalib1}.
The peak was then fitted by Gaussian and linear functions, as shown in Fig.\ \ref{Cscalib2}, and the light yield was evaluated as the mean of the Gaussian.

%\section{Analysis for observed $^{212}$Bi-$^{212}$Po like events}
\section{Identification of $^{212}$Bi--$^{212}$Po consecutive events and evaluation for number of $^{220}$Rn atoms}

\subsection{Data collection}
After introducing four liters of carrier xenon gas with $^{220}$Rn from the source, data collection was conducted for 61 h. 
The first hour of data was found to be useless because the recorded FADC range was not appropriate and rejected.

\subsection{Search for $^{212}$Bi--$^{212}$Po consecutive events}
\begin{figure}[h!]
\begin{minipage}{0.5\columnwidth}
\centering
\includegraphics[clip,width=0.9\columnwidth]{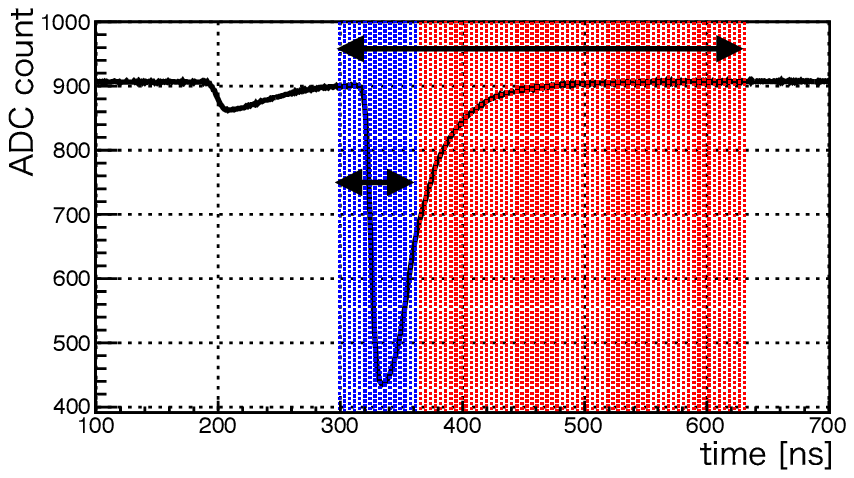}
\caption{Integration was performed only for the maximum peak; more specifically, integration within the blue timing period (with short arrow, i.e., a maximum time$-$30 ns +35 ns) and blue-and-red timing period (with long arrow, i.e., a maximum time$-$30 +300 ns) was used for pulse shape discrimination.} 
\label{BiPo_PSD}
\end{minipage}
\hspace{5mm}
\begin{minipage}{0.5\columnwidth}
\centering
\includegraphics[clip,width=1.0\columnwidth]{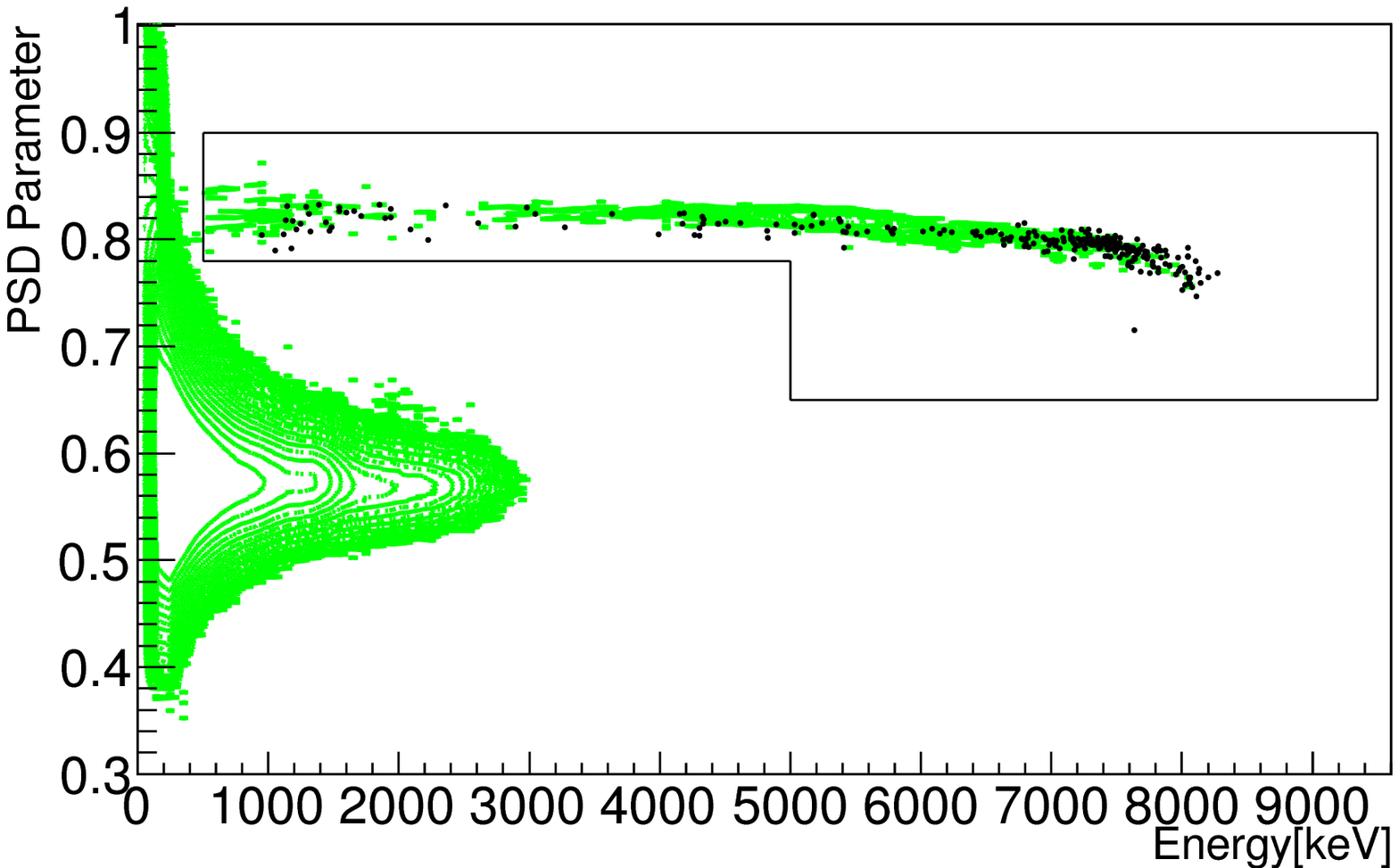}
\caption{Distribution of energy and the PSD parameter (see text), with green contours showing all events, events within the black box were selected as $\alpha$-ray events, with black dots showing $^{212}$Bi--$^{212}$Po candidates.}
\label{hPSD_BiPo}
\end{minipage}
\end{figure}

\begin{figure}[h!]
\begin{minipage}{0.5\columnwidth}
\centering
\includegraphics[clip,width=1.0\columnwidth]{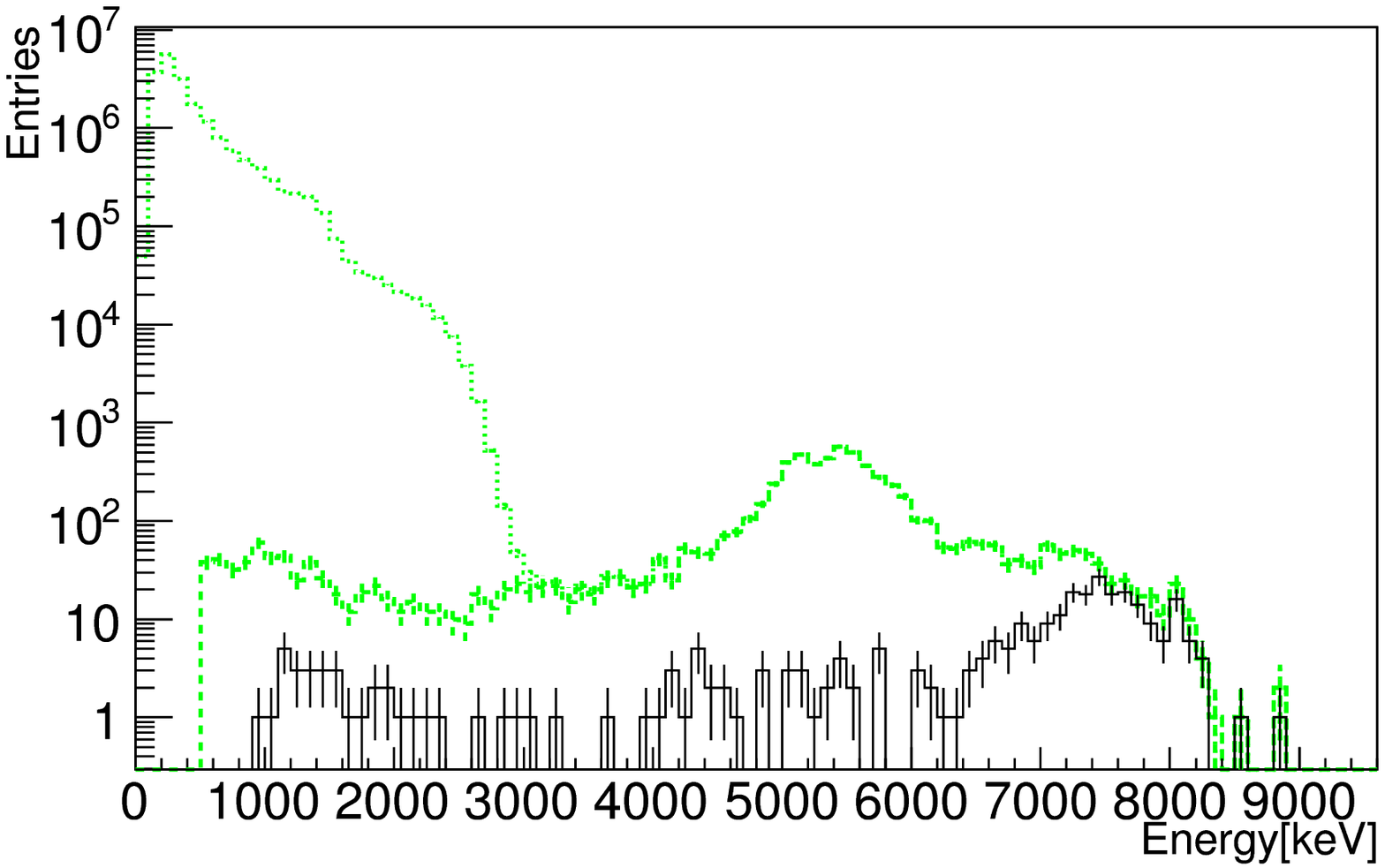}
\caption{Energy of the maximum peak of each event, with the green dotted line respecting all events, the green dashed line respecting $\alpha$-like events, and the black solid lines respecting $^{212}$Po candidates.} 
\label{E_BiPo}
\end{minipage}
\hspace{5mm}
\begin{minipage}{0.5\columnwidth}
\centering
\includegraphics[clip,width=1.0\columnwidth]{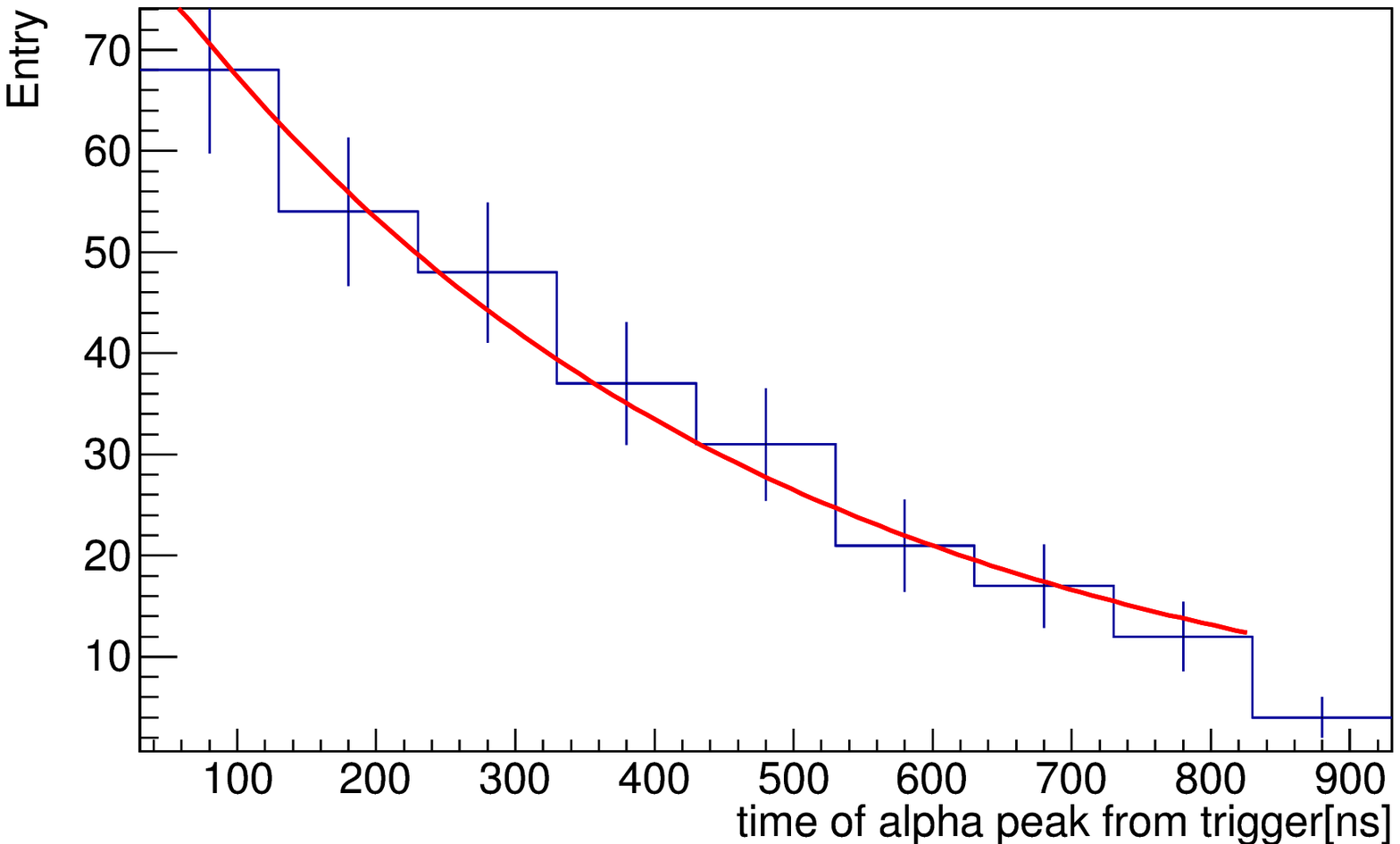}
\caption{Distribution of the timing difference between $^{212}$Bi--$^{212}$Po candidates; here, the histogram with error bars shows the observed data, while the red line shows a fitted exponential curve with the decay constant of the fitted curve observed to be consistent with that of $^{212}$Po.}
\label{BiPo_Po212}
\end{minipage}
\end{figure}

To identify $^{212}$Bi--$^{212}$Po consecutive events, we required the following two conditions:
(1) the peak with the maximum height in a waveform must be alpha-like; and 
(2) the events must have been triggered by another lower peak with the timing difference between the peak at the trigger position and peak position with the maximum height larger than 30 ns but smaller than 900 ns.
Condition (1) utilizes the fact that $\alpha$ particles have a shorter decay constant of scintillation lights than that of $\beta$ and $\gamma$-ray particles.
To discriminate $\alpha$ and $\beta$ / $\gamma$, the ratio between two integrated areas of the FADC waveform was used to provide a comparison.
The ratio of the pulse shape discrimination (PSD) parameter is defined as the blue area of Fig.\ \ref{BiPo_PSD} divided by the sum of the blue and red areas shown in the figure.
The energy of an event was obtained by summing the area observed in waveforms from the top and bottom PMTs, and calibration data was calculated via the $^{137}$Cs source, assuming linearity of the energy response.
Figure \ref{hPSD_BiPo} shows the PSD parameter versus the observed energy.
In the figure, the events within the black box were selected as alpha-like events.
Condition (2) requires the presence of a peak at the trigger position that differs from the maximum peak identified as $\alpha$-like. 
If the timing of the $\alpha$-like peak is within the range of 30 to 900 ns after the trigger timing, it was selected as the final $^{212}$Bi--$^{212}$Po consecutive events. 
As a result of our analysis here, we found 292 events in our data over 60 h. 
The expected number of events due to chance coincidence was evaluated as 0.5 and was not subtracted. 
The black dots in Fig.\ \ref{hPSD_BiPo} show the $^{212}$Po events of our final data sample, while Fig.\ \ref{E_BiPo} shows the energy distribution of the $\alpha$-ray events. 
From this data sample, a distribution of the timing differences between $^{212}$Bi and $^{212}$Po was obtained and is shown in Fig.\ \ref{BiPo_Po212}. 
The distribution was fitted by a single exponential function; 
as a result of this fit, the decay constant of $^{212}$Po candidates was 297 $\pm$ 34 ns, which is consistent with the expectation of 299 ns. 
Therefore, this supports the claim that the observed consecutive events are due to $^{212}$Bi--$^{212}$Po decays.

\subsection{Evaluation of $^{220}$Rn atoms}
There are two methods for evaluating the number of $^{220}$Rn atoms introduced. 
One method is based on the number of observed $^{212}$Bi--$^{212}$Po consecutive events, while the other method is based on the number of $\alpha$-rays due to $^{220}$Rn and $^{216}$Po just after the introduction of $^{220}$Rn.
Table 1 shows a summary of our analysis here. 
In short, $6\times10^2$ pairs of $^{212}$Bi--$^{212}$Po consecutive events were observed in this data-set though a small discrepancy between the two methods was also observed. 
A more detailed description of our two analysis follows.

\begin{figure}[h!]
\begin{minipage}{0.5\columnwidth}
\centering
\includegraphics[clip,width=1.0\columnwidth]{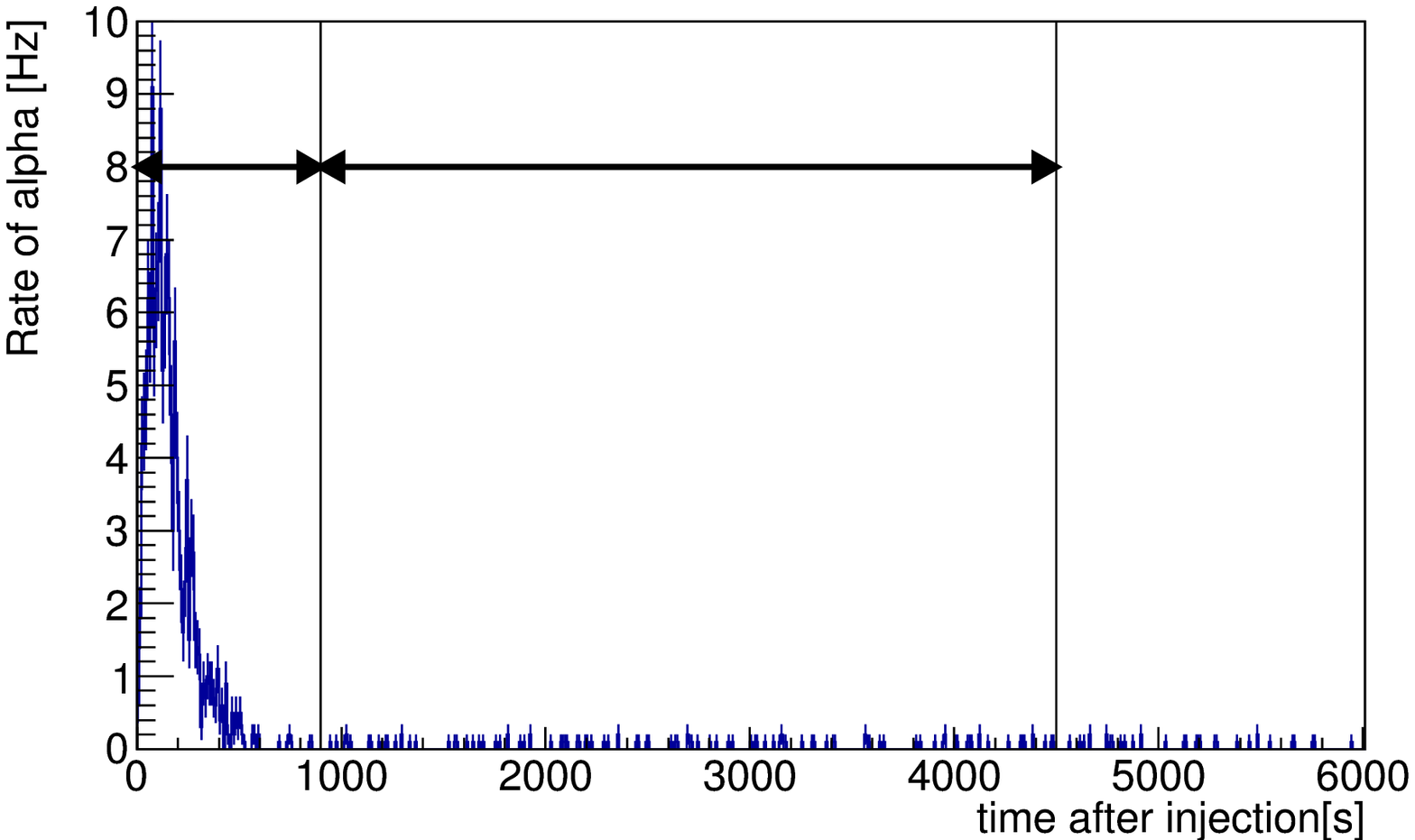}
\caption{Event rate of the injected $\alpha$-ray. Short arrow shows the time region of data, and long arrow shows that of background.} 
\label{aRate}
\end{minipage}
\hspace{5mm}
\begin{minipage}{0.5\columnwidth}
\centering
\includegraphics[clip,width=1.0\columnwidth]{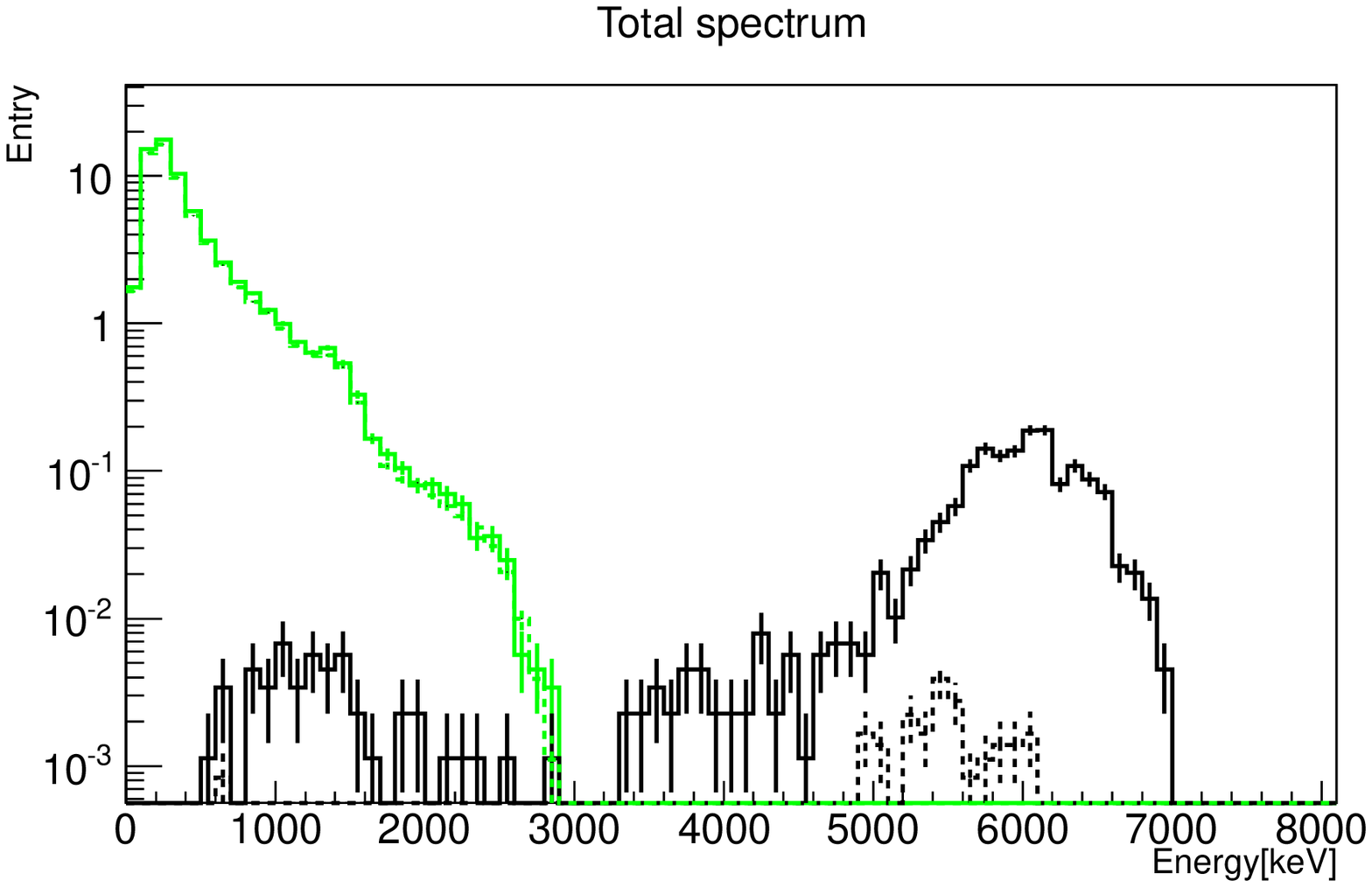}
\caption{Energy spectrum of $\alpha$ (black) and $\beta$ / $\gamma$ (green), with the solid line showing $\alpha$-rays from $^{220}$Rn and $^{216}$Po and the dotted line showing background events.} 
\label{Tnalpha}
\end{minipage}
\end{figure}

\begin{itemize}
\item The first method yielded a total of 608.3 $\pm$ 35.6 atoms after taking into account several factors, i.e., the branching ratio of the consecutive decay (0.64), the efficiency caused by the cut on the timing difference between the trigger and $\alpha$-like event (0.81), and the hour of ``dead time" just after the introduction (0.93).
\item The second method yielded a total of 700.5 $\pm$ 19.0 atoms. 
Since the event rate was high (i.e., $\sim$10 Hz, as shown in Fig.\ \ref{aRate}) just after introduction of $^{220}$Rn, tagging each $^{220}$Rn and $^{216}$Po event was not possible, because the half-life of $^{216}$Po is 0.145 seconds;
however, $\alpha$-like events were separated from $\beta$ and $\gamma$-ray events using the difference of scintillation decay constants. 
As a result, the number of observed $\alpha$-like events was 1427 in 15 min after the injection of $^{220}$Rn. 
The number of expected background atoms was evaluated to be 26.5 using data after the first hour and should therefore be subtracted from the observed number of events. 
The final number was obtained by taking into account two $\alpha$-like events observed from a single $^{220}$Rn atom. 
The energy spectrum of the observed $\alpha$-like event is shown in Fig.\ \ref{Tnalpha}.
\end{itemize}

\begin{table}[h!]
\caption{Number of observed events in each decay process and evaluated number of $^{220}$Rn atoms}
\centering
\begin{tabular}{ccc} 
\br
radioisotope & observed event & evaluated number of $^{220}$Rn atoms \\ 
\mr
$^{220}$Rn and $^{216}$Po   & 1401 & $700.5\pm19.0$ \\  
$^{212}$Bi--$^{212}$Po   & 292 & $608.3\pm35.6$ \\ 
\br
\end{tabular}
\label{N_evaluate}
\end{table}

\section{Summary}
In this paper, we presented our work in developing a $^{220}$Rn calibration source for liquid noble gas detectors. 
The key advantage of using this source is that it can provide $^{212}$Bi--$^{212}$Po consecutive events, thus enabling us to evaluate vertex resolution of a detector at low-energy by comparing low energy events of $^{212}$Bi and corresponding higher-energy $\alpha$-rays from $^{212}$Po.
Since $^{220}$Rn is a noble gas, a hot metal getter can be used when $^{220}$Rn is introduced using xenon as the carrier gas. 
In addition, no long-life radioactive isotopes are left behind in the detector after the calibration; 
this is substantially more advantageous than the method using $^{222}$Rn which leaves long-life radioactivity in the term of $^{210}$Pb. 
Using a small LXe test chamber, we developed a system to introduce $^{220}$Rn with a xenon carrier gas and demonstrated the successful introduction of $6\times10^2$ $^{220}$Rn atoms.


\section*{References}
\begin{thebibliography}{9}
\bibitem{lux} Akerib~D~S et al.  (LUX Collaboration) 2014 {\it Phys. Rev. Lett.} {\bf 112} 091303
\bibitem{XENON100}Aprile~E et al. (The XENON100 Collaboration) 2012 {\it Phys. Rev. Lett.} {\bf 109} 181301
\bibitem{DS50}Agnes~P et al. 2015 {\it Physics Letters B} {\bf 743} 456
\bibitem{xmass} Abe~K et al. (XMASS Collaboration) 2013 {\it Nucl.\ Instr.\ and Meth.\ in Phys.\ Res.\ A} {\bf 716} 78
\bibitem{NDKim} Kim~Y~D et al. (XMASS collaboration) LRT2006 http://lrt2006.in2p3.fr/talks/xmass-ydkim.pdf
\end{thebibliography}
\end{document}